\documentclass[twocolumn,showpacs,preprintnumbers,amsmath,amssymb]{revtex4}

\usepackage{graphicx}
\usepackage{dcolumn}
\usepackage{bm}

\begin{document}

\title{Electrical generation of pure spin currents in a two-dimensional electron gas}

\author{S. M. Frolov}
\author{A. Venkatesan}
\thanks{{\it Present Address}:  School of Physics \& Astronomy, University of Nottingham, Nottingham, NG72RD, U.K.  Work done while at University of British Columbia.}
\author{W. Yu}
\author{J. A. Folk}
\thanks{to whom correspondence should be addressed: {\it jfolk@physics.ubc.ca}}
\affiliation{Department of Physics and Astronomy, University of British Columbia, Vancouver, BC V6T 1Z4, Canada}
\author{W. Wegscheider}
\affiliation{Instit\"{u}t fur Angewandte und Experimentelle Physik, Universit\"{a}t Regensburg, Regensburg, Germany }

\date{\today}

\begin{abstract}
Pure spin currents are measured in micron-wide channels of GaAs two-dimensional electron gas (2DEG).  Spins are injected and detected using quantum point contacts, which become spin polarized at high magnetic field.  High sensitivity to the spin signal is achieved in a nonlocal measurement geometry, which dramatically reduces spurious signals associated with charge currents.   Measured spin relaxation lengths range from  30$\mu$m to 50$\mu$m, much longer than has been reported in GaAs 2DEG's. The technique developed here provides a flexible tool for the study of spin polarization and spin dynamics in mesoscopic structures defined in 2D semiconductor systems.
\end{abstract}

\maketitle

\noindent Interest in the physics of spin in solid state devices is driven both by the technological promise of spin electronics, and by the insights that may be gained by using spin currents as a probe into interacting electron systems.\cite{WolfScience01,ZuticRMP04}  Optical spin current measurements have advanced our understanding of spin relaxation, accumulation and separation via spin-orbit interaction in a variety of bulk semiconductors and quantum wells.\cite{KikkawaScience97,KatoScience04,CrookerScience05,SihNatphys05,SihPRL06,HolleitnerPRL06,WeberPRL07,MeierNphys07}  Spin currents can also be generated and detected electrically using spin-selective contacts, enabling straightforward integration into circuits where device geometry and spin parameters are controlled by gates.\cite{JohnsonPRL85,JedemaNature01,PotokPRL02,FolkScience03,RokhinsonPRL04,ValenzuelaNature06,TombrosNature07, KoopExp08}

Devices defined by electrostatic gates in GaAs/AlGaAs two-dimensional electron gases (2DEG's) display an extraordinary variety of spin-related phenomena, showing promise for quantum dot-based quantum information processing, coherent spin rotations mediated by spin-orbit interaction, even the possibility of spontaneous spin polarization in quantum point contacts.\cite{NowackScience07,ThomasPRL96}  These structures are typically studied using direct measurements of the charge currents passing through them.\cite{FolkScience03, HansonPRB04, RokhinsonPRL04}  The sensitivity to spin properties can be greatly enhanced by measuring pure spin currents resulting from spin-resolved charge transport, but such measurements have not yet been integrated with gate-defined mesoscopic devices.\cite{JohnsonPRL85}

In this Letter, we present electrical measurements of pure spin currents in micron-wide channels of a GaAs 2DEG using one-dimensional constrictions known as quantum point contacts (QPC's) as injectors and detectors.\cite{vanweesprl88,WharamJPCSSP88}  The ability to change the channel geometry in-situ using gate voltages enabled an accurate measurement of spin relaxation length even for small contact polarizations.  The relaxation lengths observed in this work, $\lambda_s=30-50\mu m$, are significantly longer than the values typically reported in GaAs 2DEG's because spin-orbit mediated relaxation was suppressed by the external magnetic field.\cite{MillerPRL03,Ivchenko73, DuckheimPRB07}  The temperature- and field-dependences of the spin current polarization were used to extract a Lande g-factor in the QPC's, $|g|=0.75\pm0.1$, that is enhanced compared to $|g|=0.44$ in the bulk.\cite{ThomasPRL96}  An advantage of this polarization-based g-factor measurement is that it does not depend on the interpretation of QPC conductance features.

Pure spin currents are generated electrically through a sequence of two processes.  First, charge is injected across a spin-selective barrier, creating a higher population of one spin. Next, the nonequilibrium spin population that accumulates outside of the injector diffuses towards a large electrically floating reservoir with spins in equilibrium.  Experimental realizations of this technique often rely on ferromagnetic contacts,\cite{JohnsonPRL85,JedemaNature01,ValenzuelaNature06,LouNphys07} but injection from ferromagnets into GaAs 2DEG's remains a challenge. QPC's in Tesla-scale magnetic fields are a natural alternative because they are defined within the 2DEG itself.\cite{vanweesprl88,WharamJPCSSP88}  In contrast to ferromagnets, the polarization axis of a QPC is aligned with the external magnetic field so no Hanle precession of spin currents is expected.

Figure 1(a) shows a schematic of the measurement. A voltage, $V_{ac}$, is applied across a spin-selective injector QPC driving polarized current, $I_{inj}$, into the center of a long channel.  The spin population that accumulates above the injector diffuses toward the large 2DEG reservoirs at the left and right ends of the channel. All charge current flows to the electrical ground at the left end of the channel;  pure spin current flows to the right.  The detector QPC, located a distance $x_{id}$ to the right of the injector, measures the nonlocal voltage, $V_{nl}$, due to spin accumulation generated by the pure spin current.

\begin{figure}[t]
\includegraphics[width=0.5\textwidth]{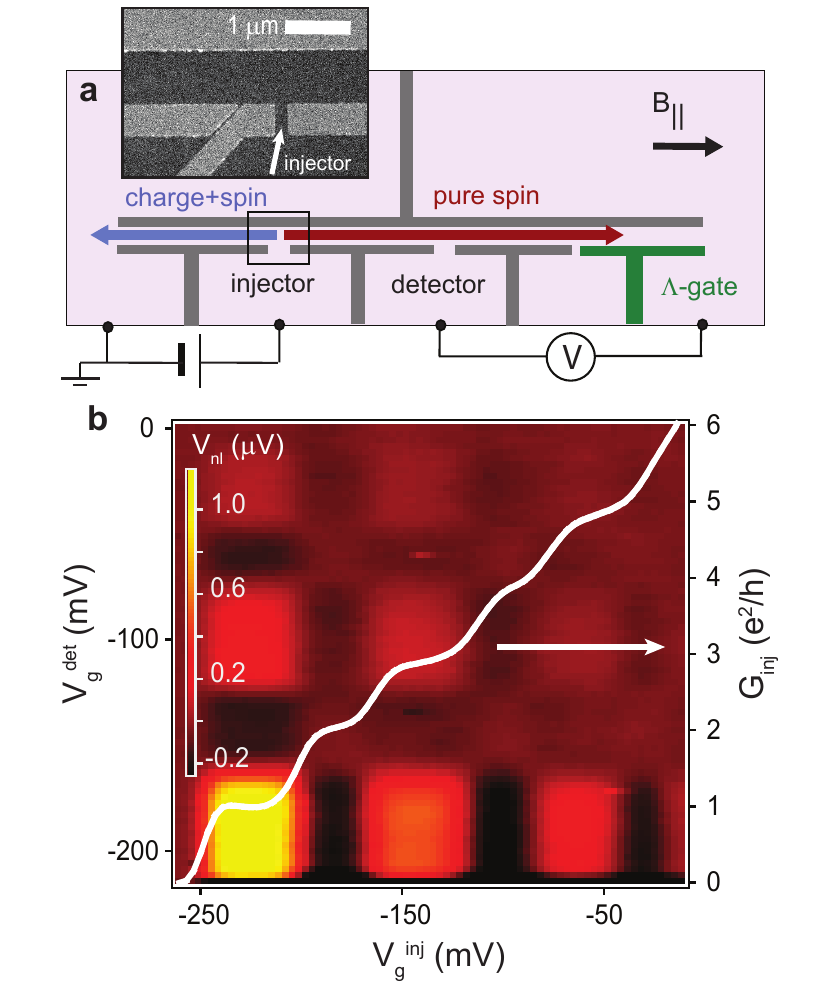}
\caption{  (a) Schematic of nonlocal measurement geometry.  Depleted gate pattern (dark gray) simplified for clarity. Inset: scanning electron micrograph (SEM) of typical QPC (gates are light gray in SEM image).  Nonlocal voltages reported in this paper are for the detector region with respect to the right reservoir.  (b) Nonlocal voltage as injector (bottom axis) and detector (left axis) QPC's are scanned through polarized and unpolarized settings using $V_g$ ($B_{||}=10T$, $T=500mK$, $V_{ac}=50\mu V$, $x_{id}=6.7\mu m$). Injector conductance shown in white (right axis).  Relative magnitudes of the signal at different spin-polarized squares reflect reduced polarization at higher odd QPC plateaus ($G=3e^2/h,5e^2/h, ...$), partially counteracted by higher injector currents in a voltage-biased configuration.\label{DeviceGeometry}}
\end{figure}

The devices were defined using electrostatic gates on the surface of a [001] GaAs/AlGaAs heterostructure. The 2DEG was 110 nm below the surface, with electron density $n_s=1.11\times 10^{11}cm^{-2}$ and mobility $\mu=4.44\times 10^{6} cm^2/Vs$ measured at $T=1.5K$.  The data in this paper are from three channels, each along the [110] crystal axis, with lithographic width $1\mu m$ and length  $100\mu m$.   The injector and detector spacing ranged from $x_{id}=3-20\mu m$.  Lock-in measurements in a dilution refrigerator were performed in magnetic fields, $B_{||}$, applied along the channel axis. To avoid trajectories dominated by skipping orbits, the out-of-plane component, $B_\perp$, was kept under $50mT$, ensuring that the cyclotron radius was greater than the channel width. The effective sheet resistance in the channel, $\rho_\square \sim 20-120\Omega$, depended on cooldown conditions.   The resistance increased by 10-20\% from $B_{||}=0$ to $B_{||}=10T$.

Gate voltages control QPC conductance, $G(V_g)$, and polarization, $P(V_g)$.   $G(V_g)$ is quantized in units of  $1e^2/h$ at high magnetic field, as spin-resolved one-dimensional subbands are added one by one.  The first ($G=1e^2/h$) plateau corresponds to fully polarized transmission, $P=(G_\uparrow-G_\downarrow)/(G_\uparrow+G_\downarrow) \sim 1$, as only a single spin-up subband is allowed through the QPC ($G_\downarrow \sim 0$).  The second ($G=2e^2/h$) plateau  corresponds to unpolarized transmission, $P=0$ (one spin-up and one spin-down subband); the third corresponds to $P=1/3$ (two spin-up and one spin-down subband), etc.

\begin{figure}[t]
\includegraphics[width=0.5\textwidth]{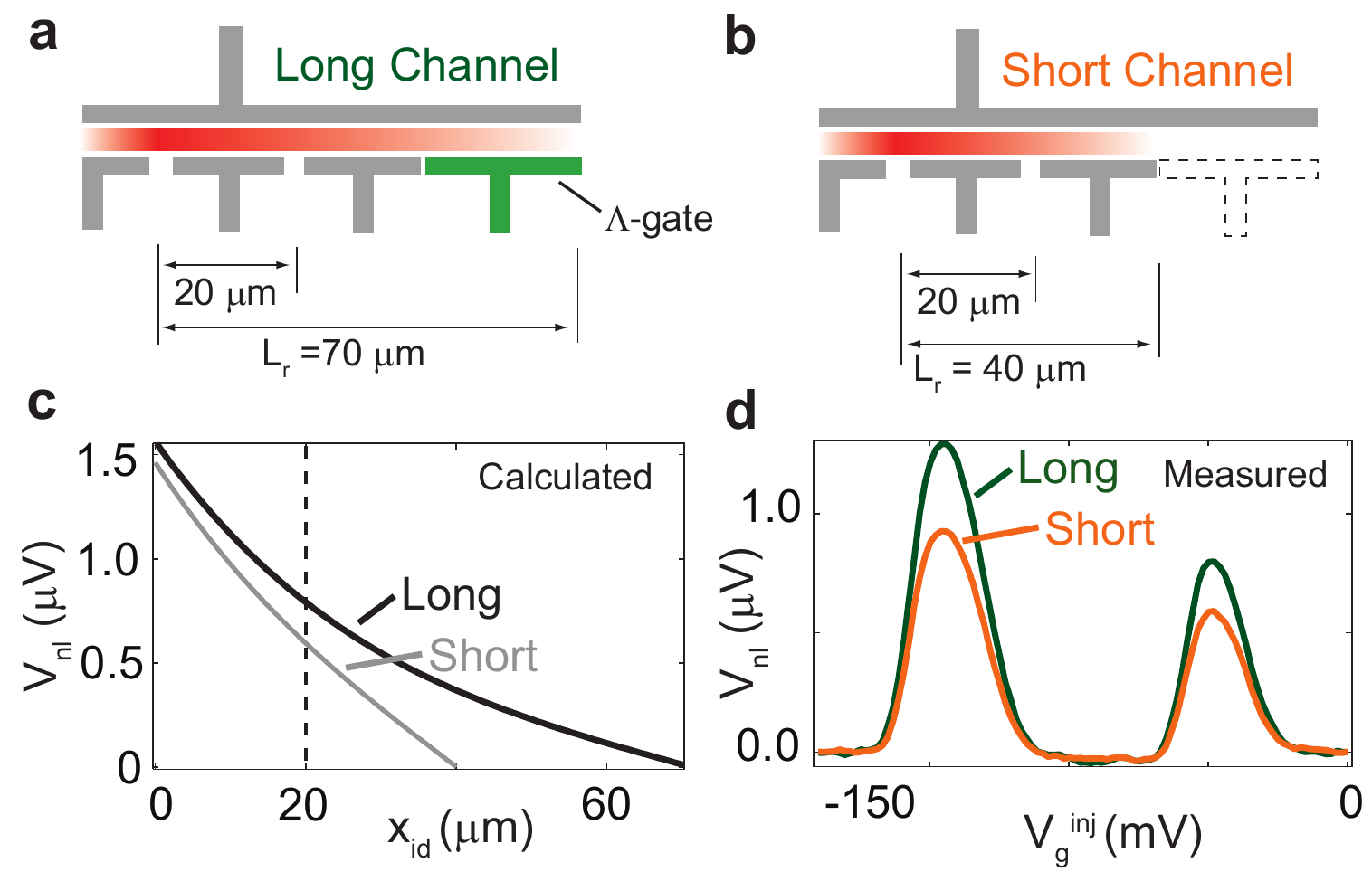}
\caption{  (a) and (b) nonequilibrium spin-up chemical potential (gradient) in the 2DEG channel with $\Lambda$-gate depleted (long) and undepleted (short). (c) Nonlocal voltage calculated from Eq.1 for a range of injector-detector spacings, using $\lambda_s=30\mu m$ and the channel lengths in panels (a) and (b). (d) Nonlocal signal measured with $\Lambda$-gate depleted and undepleted, for a device with $x_{id}=20\mu m$ and the geometry of panels (a) and (b), using a polarized detector at T=1.2K.\label{lambdagate}}
\end{figure}

Nonlocal signals measured at high magnetic field had a characteristic signature of spin currents, see Fig.1(b).   Positive voltages indicating a non-equilibrium spin population above the detector were observed when both contacts were spin-selective, i.e. when both were tuned to odd conductance plateaus ($G_{inj}$, $G_{det}$ = $1e^2/h$, $3e^2/h$, etc.).     The voltage was near zero when both the detector and the injector were set to even plateaus ($G_{inj}$, $G_{det}$ = $2e^2/h$, $4e^2/h$, etc.).  A small negative voltage was often observed when only the injector or only the detector was polarized (e.g., [$G_{inj}=2e^2/h$, $G_{det}=1e^2/h$] or [$G_{inj}=1e^2/h$, $G_{det}=2e^2/h$]).  The origins of the negative signal are currently under investigation.

The spin signal depends in general on a combination of diffusion, spin relaxation, and contact polarization.  The expected magnitude of the nonlocal voltage can be calculated from a 1D diffusion equation with boundary conditions of equilibrium polarization at the left and right ends of the channel (distances $L_l$ and $L_r$ from the injector), and including spin relaxation characterized by length $\lambda_s$:

\begin{equation}
V_{nl}(x_{id}) =  \frac{\rho_{\square} \frac{\lambda_s}{w} I_{inj} P_{inj} P_{det} sinh(\frac{L_r - x_{id}}{\lambda_s})}{sinh(L_r/\lambda_s) (coth(L_r/\lambda_s)+coth(L_l/\lambda_s))}
\end{equation}

\noindent where $w$ is the channel width.

One way to measure spin relaxation length is to compare $V_{nl}$ across several detectors at different positions along the channel.  But this technique relies on identical detector polarizations---not necessarily the case for QPC's at intermediate values of $B_{||}$ and finite temperature.  The flexibility of the gate-defined geometry enabled a measurement of spin relaxation length that was independent of $P_{inj}$ and $P_{det}$.

The bottom wall of the channel to the right of the detector was defined by two gates, see Fig.1(a).  Undepleting the $\Lambda$-gate shortens the right side of the channel, bringing the right-hand equilibrium spin reservoir closer to the detector (Figs.2(a),(b)) and causing a faster drop in the spin-up chemical potential along the channel (Fig.2(c)).  If the spin current has relaxed before reaching the $\Lambda$-gate, the effect of undepleting the $\Lambda$-gate is negligible.  But for a channel with $\lambda_s \gtrsim L_r$, the nonlocal signal decreases when the $\Lambda$-gate is undepleted (Fig.2(d)), and $\lambda_s$ can be extracted from the ratio of the signals for long and short channels using Eq.(1).  Different channels and different cooldowns gave values of $\lambda_s$ that ranged from $30\mu m$ to $50\mu m$, and were independent of field and temperature from $B_{||}=3-10T$ and $T=50mK-2K$.

The primary cause of spin relaxation in high-mobility GaAs 2DEG's is a trajectory-dependent effective magnetic field, $B_{so}$, arising from spin-orbit interaction.\cite{DyakonovSPSS72}    Spin relaxation by this mechanism is, in general, suppressed in a large external magnetic field, $B_{||}\gg B_{so}$.\cite{Ivchenko73, DuckheimPRB07}   Monte Carlo simulations of spin dynamics due to a spin-orbit field were made using the channel geometry from this work and considering a range of spin-orbit parameters.\cite{Luescher}  [110]-oriented spins relax due to the component of $B_{so}$ along the [\={1}10] axis; the simulations suggest  an upper limit $B_{so}$[\={1}10]$<$1.5T in order to find $\lambda_{s}>30\mu m$ over the field range $B_{||}=3-10T$.  In contrast to the experimental results, the simulations also show $\lambda_{s}$ to be strongly dependent on the external field, rising to greater than $300\mu m$ at $B_{||}=10T$.  Other spin relaxation mechanisms may limit the measured $\lambda_{s}$ and account for the discrepancy.\cite{ElliottPR54}

\begin{figure}[b]
\includegraphics[width=0.5\textwidth]{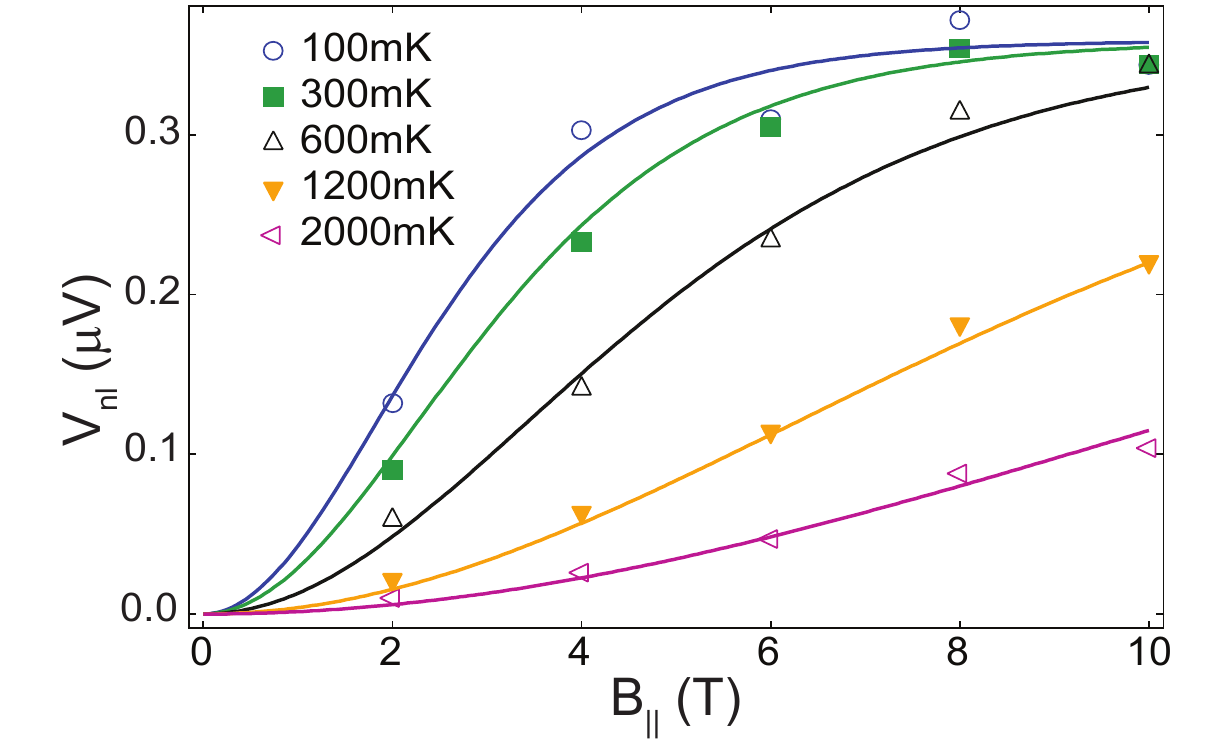}
\caption{ Peak nonlocal signal for $G_{inj}$ and $G_{det}$ near $1e^2/h$, across a range of magnetic fields and temperatures.  All data are from a single cooldown, with $V_{ac}=10\mu V$ and $\rho_\square\sim120\Omega$.  Solid lines show fit of QPC model to data.  Fits do not include data at zero field because the distinction between thermoelectric and spin signals was ambiguous (see text).\label{fielddep}}
\end{figure}

Spin current measurements can be used to quantify spin-selective transmission of the injector and detector.  A simple model of a QPC is a saddle point potential barrier that couples two leads with thermally-broadened Fermi distributions  and Zeeman-split spin populations.  In general, QPC polarization approaches $P=1$ when Zeeman energy $g\mu_B B$ is much larger than both thermal broadening $k_B T$ and tunnel broadening $\hbar \omega$.    Polarization results from different spin-resolved conductances: $G_{\uparrow [\downarrow]}(E_0) = \int \frac{df(E+[-] \frac{g\mu_B B_{||}}{2}, T)}{dE} T(E-E_0) dE$, with subband cutoff energy $E_0(V_g)$ and transmission $T(E)=1/(1+e^{-2\pi E/\hbar \omega})$.  The evolution of the spin signal in magnetic field and temperature (Fig.3) is consistent with a constant relaxation length and QPC polarization that would be expected from the saddle point model with g-factor $|g|=0.75\pm 0.1$ and tunnel broadening $\hbar \omega=190 \pm 20 \mu eV$ (Fig.3).    Similar g-factors were found for all devices.  Enhanced g-factors extracted from conductance signatures (rather than QPC polarization) have previously been ascribed to stronger exchange interaction at low density.\cite{ThomasPRL96}

\begin{figure}[t]
\includegraphics[width=0.5\textwidth]{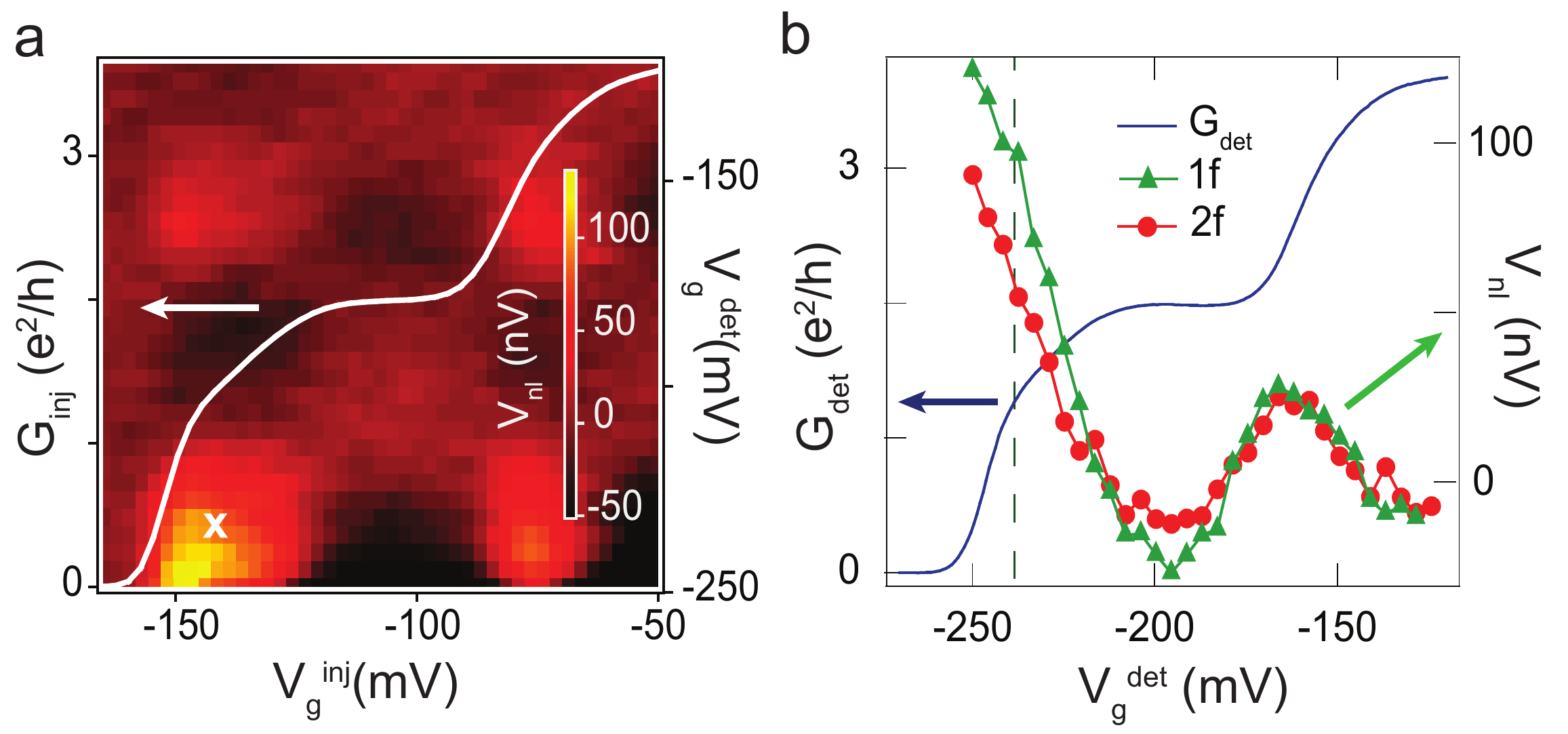}
\caption{(a) Colorscale: first harmonic of the nonlocal signal at $B_{||}=0$. White trace shows conductance of injector QPC (left axis).  Gate settings used to estimate Peltier coefficient indicated with ``x". (b) $1f$ and $2f$ nonlocal signals (right axis) correlate with conductance of detector QPC (left axis), measured with injector QPC at ``x" from (a). Dashed line indicates detector gate setting used to calibrate thermoelectric sensitivity. ($V_{ac}=50\mu V$, $T=500mK$, $B_{||}=0$)} \label{Zerofield}
\end{figure}

Nonlocal voltages unrelated to spin accumulation were also observed.  Fluctuations due to quantum interference were superimposed on the spin signal, but were within experimental noise for $x_{id} > 10\mu m$ or $T > 200mK$.\cite{SkocpolPRL87}   Joule and Peltier heating of the channel by the injected current gave rise to a temperature difference across the detector, $\Delta T$, that led to thermoelectric contributions to the nonlocal voltage.\cite{MolenkampPRL90,MolenkampPRL92}  Signals due to Joule heating did not interfere with the spin signal because they appeared at the second harmonic ($2f$) of the lock-in excitation, $\Delta V_{Joule}=S_{det} \Delta T \propto S_{det}I_{inj}^2$, where $S_{det}$ is the thermopower of the detector QPC.

In contrast to Joule heating, Peltier heating appears at the first harmonic ($1f$) of the excitation: $\Delta V_{Peltier} \propto S_{det}S_{inj} T I_{inj}$, and was more difficult to distinguish from the spin signal.  An identifying characteristic of the spin signal was its magnetic field dependence: the spin component was significantly larger than the thermoelectric voltage for $B_{||}>3T$, but the distinction was ambiguous at lower fields.  A nonlocal signal that remained clearly visible down to zero field in the experiment motivated a more careful analysis of the thermoelectric contribution.

Figure 4(a) illustrates the similarity between spin and thermal signatures at low magnetic field (cf. Fig.1(b)).  QPC thermopower is zero on conductance plateaus, but finite at the transitions between plateaus as well as on the so-called 0.7 structure that is commonly observed at low field.\cite{MolenkampPRL90,ThomasPRL96,AppleyardPRB02}   Finite thermopower for injector and detector near the steps in conductance gives rise to a Peltier signal in a checkerboard pattern that is reminiscent of the spin signal.  The thermoelectric origin of the $1f$ signal in Fig.4(a) is supported by a comparison of the zero-field signals at $1f$ and $2f$ (Fig.4(b)). The $2f$ signal is proportional to Joule heating by the injected current and to the thermopower of the detector, and serves as a fingerprint of thermal effects.  The $1f$ signal shows a nearly identical gate voltage dependence to the $2f$ signal, suggesting that it is also thermal.  The $2f$ signal can be used to extract the thermoelectric sensitivity of the detector QPC to heating: $V_{nl}/(I^2R)=1\pm0.1nV/fW$ at the first detector conductance step. Assuming that the $1f$ signal is due entirely to Peltier heating through the injector, the magnitude of the signal at the first injector and detector conductance steps implies $S_{inj}=100\pm10 \mu V/K$ at $T=500mK$, consistent with previous measurements.\cite{MolenkampPRL90,MolenkampPRL92, AppleyardPRL98}

Spin selectivity of QPC's at zero magnetic field has been linked to 0.7 structure in earlier experiments.\cite{ThomasPRL96, RokhinsonPRL06}  The analysis above shows that the data in Fig.4(a) may be explained without invoking a spontaneous spin polarization.   It does not rule out a small additional contribution due to spin, but no direct evidence for zero-field spin polarization was observed.  For example, Hanle precession due to milliTesla-scale external fields would have been expected if the polarization axes of the QPC's were fixed by an intrinsic broken symmetry.  If the polarization axes were not fixed, an increase in the signal might have been expected as uncorrelated axes were aligned by a small external magnetic field.  To look for these effects, small fields were applied along [110] and [\=110], but no change in the signal was observed up to several hundred milliTesla, where conventional QPC polarization sets in.

In conclusion, pure spin currents travel for tens of microns in micron-wide channels of 2DEG, and provide a valuable probe of spin relaxation and spin polarization in mesoscopic structures.  In the future, spin-orbit anisotropy in GaAs 2DEGs will be explored by rotating the channel axis and the direction of the external in-plane magnetic field.

{\small {\bf Acknowledgements:}  The authors thank M.~Duckheim, J.C.~Egues, D.~Loss, S.~L\"uscher, and G.~Usaj for valuable discussions. Work at UBC supported by NSERC, CFI, and CIFAR.  W.W. acknowledges financial support by the Deutsche Forschungsgemeinschaft (DFG) in the framework of the program  ``Halbleiter-Spintronik'' (SPP 1285).}

{\small

\bibliographystyle{apsrev}
\bibliography{spindiff}}

\end{document}